\documentclass[prl,showpacs,amsmath,amssymb,letter,preprint,endfloats]{revtex4}
\usepackage{graphicx}
\begin{document}
\title{Precessional dynamics of elemental moments\\ in a ferromagnetic alloy}
\author{W.E. Bailey and L. Cheng}
\affiliation{Dept. of Applied Physics, Columbia University, 500 W
120th St, New York, NY 10027}
\homepage{http://magnet.ap.columbia.edu}
\author{D. J. Keavney}
\affiliation{Advanced Photon Source, Argonne National Laboratory,
Argonne IL 60439}
\author{C.-C. Kao, E. Vescovo, and D.A. Arena}
\affiliation{National Synchrotron Light Source, Brookhaven
National Laboratory, Upton NY}
\date{15 Mar. 2004}

\begin{abstract} We demonstrate an element-specific measurement of
magnetization precession in a metallic ferromagnetic alloy,
separating Ni and Fe moment motion in Ni$_{81}$Fe$_{19}$.
Pump-probe X-ray magnetic circular dichroism (XMCD), synchronized
with short magnetic field pulses, is used to measure free
magnetization oscillations up to 2.6 GHz with elemental
specificity and a rotational resolution of $<$ 2 $^{\circ}$.
Magnetic moments residing on Ni sites and Fe sites in a
Ni$_{81}$Fe$_{19}$(50nm) thin film are found to precess together
at all frequencies, coupled in phase within instrumental
resolution of 90 ps.

\end{abstract}
\maketitle
\section{Introduction}

The precession of magnetic moments in an applied magnetic field is
relevant for many classes of materials studies.  Precession is
usually observed through microwave absorption, as in electron spin
resonance (ESR), nuclear magnetic resonance (NMR), or
ferromagnetic resonance (FMR), combining the response from all
species present.  In complex materials and heterostructures,
time-domain optical techniques have proven useful to separate
precessional dynamics at different atomic sites through the photon
energy dependence of the response. Site-specific measurement of
ESR\cite{awsch-malaj} and NMR\cite{marohn} precession has been
achieved in semiconductors using visible optical pump-probe
techniques.  FMR precession of ferromagnetic alloys and
heterostructures is technologically important since it determines
data rates (switching speeds) of spin electronics devices.

The magnetic moments on $A$ and $B$ sites in a random
ferromagnetic alloy $A_{1-x}B_{x}$ are usually believed to be
parallel and collinear, due to strong exchange forces, either in
static equilibrium or precessional motion. However, noncollinear
static moments are predicted as the ground state in random
Ni$_{1-x}$Fe$_{x}$ alloy systems, particularly near
$x=0.35$.\cite{van-s-nature}  The random site disorder plays a
strong role in the reduction of collinearity, evidenced in the
30\% reduction of exchange stiffness for $x=0.75$ (Ni$_{3}$Fe)
upon an order-disorder transformation.\cite{j-p-f-nife-ordering}
In Ni$_{81}$Fe$_{19}$ (permalloy,) total energy calculations
predict a 1-2 $^{\circ}$ average angle of moment noncollinearity
at T=0.\cite{pc-schul}

The magnetization dynamics of Ni$_{81}$Fe$_{19}$ have been studied
more intensively than those of any other alloy. Given the
noncollinear ground-state alignment predicted for
Ni$_{81}$Fe$_{19}$, it is natural to ask whether a small net
misalignment of Ni and Fe moments may also exist in a low-energy
configuration accessible at short times and finite temperatures.
The different moments in the alloy (2.5 vs. 1.6
$\mu_{B}$/atom)\cite{krill} and different effective gyromagnetic
ratios $g_{eff}$ in elemental crystals (2.20 vs.
2.11)\cite{farle-rpp} of Ni and Fe suggest different natural
angular velocities for the motion of elemental moments. An
element-specific measurement of magnetization precession can
provide a direct test for the presence of a phase lag between
$\mathbf{M}_{Ni}$ and $\mathbf{M}_{Fe}$ in coupled motion.

An element-specific measurement of ferromagnetic resonance (FMR)
precession has not been available previously.  An x-ray optical
technique, x-ray magnetic circular dichroism (XMCD), is a
well-established, quantitative, element-specific measurement of
magnetic moment\cite{chen}; however, its previously demonstrated
time-domain extension shows resolution of $\sim\textrm{2ns}$. This
has been sufficient to characterize 180$^{\circ}$ reversal
dynamics in buried magnetic layers\cite{fontaine-prl}, but is
insufficient to characterize FMR precession at frequencies of
$f_{p}\geq\textrm{1 GHz}$.

\begin{figure}[htb]
\includegraphics[width=\columnwidth]{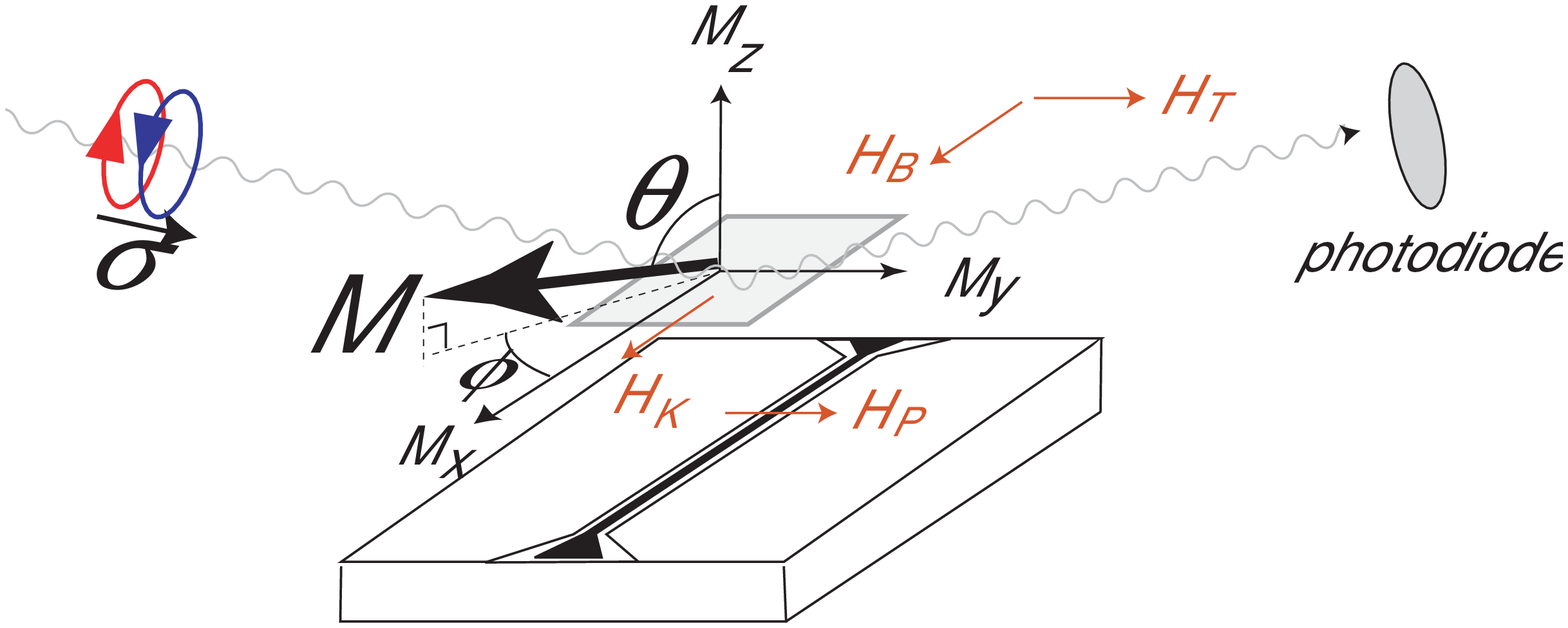}
\caption{TR-XMCD measurement configuration.  Circularly polarized
photons are reflected at grazing incidence into a photodiode
(right).  XMCD is measured by switching photon helicity
$\mathbf{\sigma}$.  Relative alignment of applied fields $H_{P}$
(pulse), $H_{B}$ (longitudinal bias), $H_{K}$ (anisotropy), and
$H_{T}$ (transverse bias) is shown. \label{diag}}
\end{figure}

We present direct experimental separation of Ni and Fe magnetic
moment precession in ferromagnetic Ni$_{81}$Fe$_{19}$(50nm)
through pump-probe XMCD.  Improved time resolution
($\sim\textrm{90 ps}$ FWHM) has been achieved in part through the
use of a fast-falltime squarewave magnetic field pulse delivered
through a coplanar waveguide (CPW).\cite{kos-pimm}  We verify
coupled precession of elemental moments within instrumental
resolution.

\section{Experimental}
Time-resolved XMCD (TR-XMCD) measurements were carried out at
Beamline 4-ID-C of the Advanced Photon Source in Argonne, IL.  The
circular dichroism signal was obtained in reflectivity at near
grazing incidence (see Fig \ref{diag}), using photon helicity
$\mathbf{\sigma}$ switching
($\mathbf{\sigma}\parallel\mathbf{\hat{y}}$) at the elliptical
undulator for fixed applied field $\mathbf{H}$. Reflected
intensity was read at a soft x-ray photodiode and normalized to an
incident intensity at a reference grid.  The measurement technique
is photon-in/photon out and should therefore minimize artifacts
from time-varying stray magnetic fields.

Time resolution was achieved through a pump-probe technique.  Fast
fall-time magnetic field pulses (pump) were synchronized with
variable delay to x-ray photon bunches from the APS storage ring
(probe.)  The repetition frequencies of photon bunches and
magnetic field pulses were both 88.0 MHz (11.37 ns period);
therefore, source or detector gating was not required.

The magnetic field delivery configuration is similar to that used
in a pulsed inductive microwave permeameter (PIMM).
\cite{kos-pimm,bailey-apl2003}.  Fast fall-time current pulses
($\sim$ 150ps) were delivered from a commercial pulse generator
and through a CPW located under the magnetic thin film, providing
pulsed transverse fields $<$ 10 Oe in amplitude.  Pulses
terminated into a 20 GHz sampling oscilloscope with 27 dB
attenuation for pulse waveform characterization or directly into a
50$\Omega$ load during TR-XMCD measurement.  The CPW center
conductor was aligned along $\mathbf{\hat{x}}$, generating pulse
fields $\mathbf{H_{P}}=H_{P}\:\mathbf{\hat{y}}$. Orthogonal
Helmholtz coils apply longitudinal bias
$\mathbf{H_{B}}=H_{B}\:\mathbf{\hat{x}}$ or transverse magnetic
field bias $\mathbf{H_{T}}=H_{T}\:\mathbf{\hat{y}}$.

Experimental system time resolution can be estimated by using the
finite photon bunch length $\sigma_{ph}$ and the timing jitter in
the current pulse delivery electronics $\sigma_{tj}$. Separate
streak camera measurements provide RMS bunch length
$\sigma_{ph}=\textrm{25 ps}$\cite{pc-lump} and sampling
oscilloscope measurements provide RMS timing jitter $\sigma
_{tj}=\textrm{30 ps}$. Adding these contributions in quadrature
yields a net timing resolution of
$\sqrt{\sigma_{ph}^{2}+\sigma_{tj}^{2}}=\textrm{39 ps}$ RMS or 90
ps FWHM.

The Ni$_{81}$Fe$_{19}$(50nm)/Ta(2nm) thin film was deposited on
the CPW using ion beam sputtering at a base pressure of
5$\times$10$^{-8}$ Torr.  Magnetic anisotropy was induced using a
static magnetic field applied along the center conductor, creating
an effective uniaxial anisotropy field
$\mathbf{H_{K}}\parallel\mathbf{\hat{x}}$. The sample normal
points along $\mathbf{\hat{z}}$; the plane of incidence of the
beam is $yz$, oriented roughly 5$^{\circ}$ off $\mathbf{y}$.
During TR-XMCD, magnetization was longitudinally biased,
$\textrm{0}<H_{B}<\textrm{100 Oe}$, along the easy anisotropy
axis, therefore $\mathbf{M}\bot \mathbf{\sigma}$, and
magnetization was rotated over small angles by the pulsed oersted
field $H_{P}\sim 7\:\mathbf{Oe}$.  Measurements were carried out
at room temperature (295 K).

\begin{figure}[htb]
\includegraphics[width=\columnwidth]{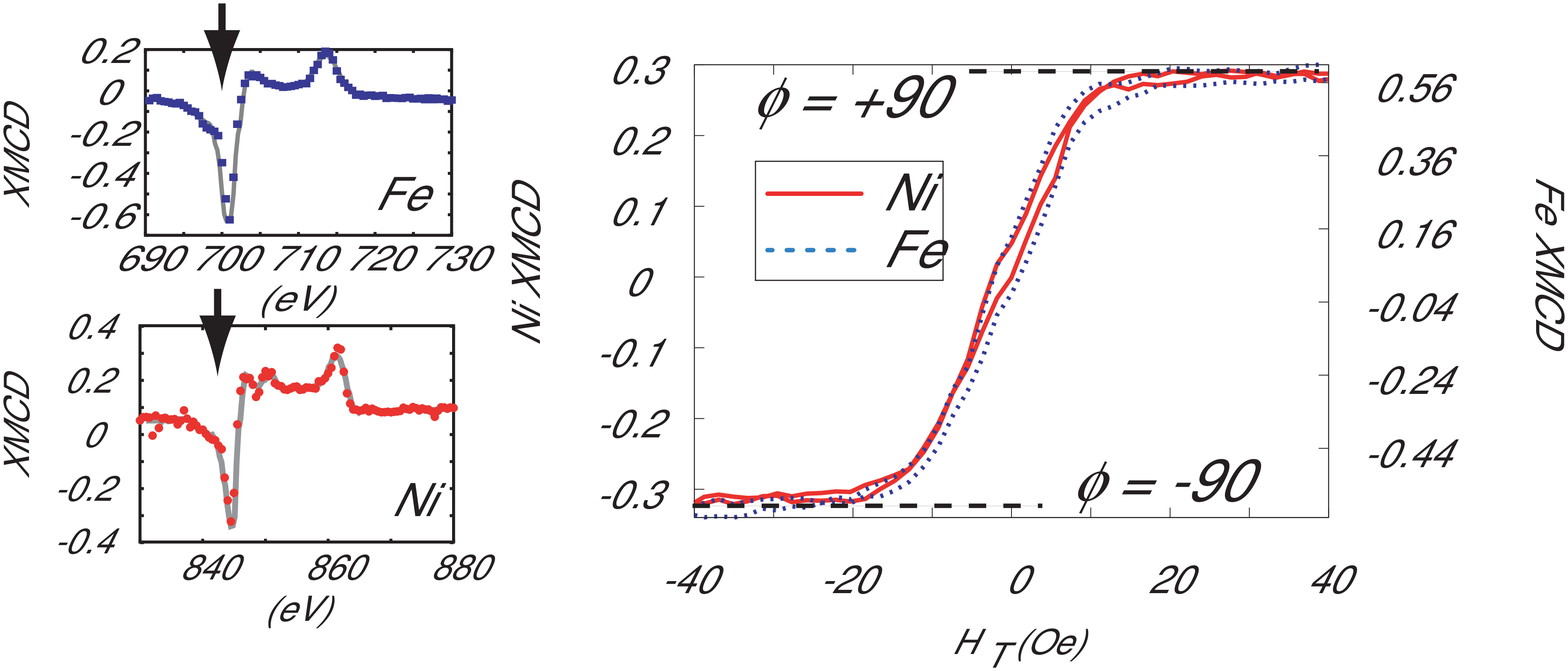}
\caption{{\it Left:} XMCD spectra over L$_{3}$ and L$_{2}$ edges
for Ni and Fe.  Photon energy was set to the $L_{3}$ peak for Ni
and Fe before measurement of element-specific hysteresis loops
({\it right.}).  The signal provides a calibration for the
elemental magnetization angles $\phi_{Ni}$ and $\phi_{Fe}$.
\label{calib}}
\end{figure}

Element-specific XMCD hysteresis loops were taken as a function of
transverse bias $H_{T}$ to obtain a calibration for rotation angle
$\phi$ (Fig \ref{calib}.) Photon energies were set to the L$_{3}$
peaks for Fe (701 eV) and Ni (844eV) to measure Fe and Ni XMCD
signals respectively (Fig \ref{calib}, left;) nominal energies
were close to reported peak energies of 707 eV and 853 eV,
respectively.  The saturation values of XMCD signals are taken to
be $\phi_{Ni}=\phi_{Fe}=\pm\textrm{90}^{\circ}$.  Well-defined
hard-axis loop behavior is shown with an anisotropy field $H_{K}$
of 11.8 $\pm$ 1.0 Oe (Fig \ref{calib}, right); the Fe dichroism
signal is roughly a factor of two larger than the Ni dichroism
signal, consistent with previous work\cite{brooks-dhole} and the
smaller number of holes in Ni.

\section{Results}

\begin{figure}[htb]
\centering
\includegraphics[width=\columnwidth]{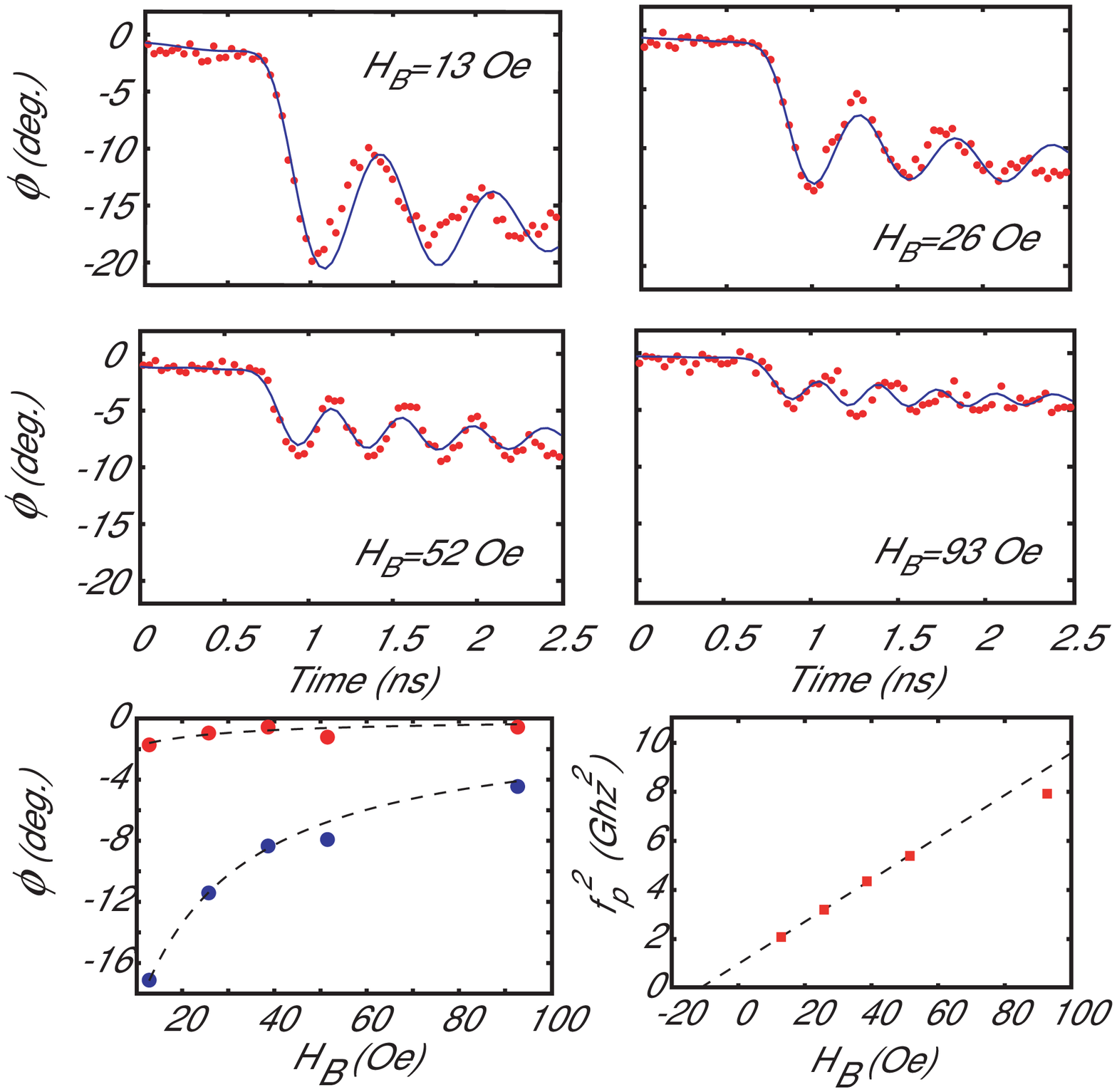}
\caption{Time-dependence of Ni magnetization angle $\phi_{Ni}(t)$
after falling magnetic field (points), with LLG simulation
(lines)\label{llg}. {\it Top left:} H$_{B}$=13 Oe. {\it Top
right:} H$_{B}$=26 Oe. {\it Middle left:} H$_{B}$=52 Oe. {\it
Middle right:} H$_{B}$=93 Oe.  {\it Bottom left:} equilibrium
magnetization angles for high and low pulse levels; {\it bottom
right:} Kittel plot of extracted precessional
frequencies.}\end{figure}

The measurement of magnetization precession for Ni moments alone
is presented in Figure \ref{llg}.  XMCD signals were measured as a
function of pump-probe delay and converted to time-dependent
elemental Ni magnetization angles $\phi_{Ni}(t)$ according to Fig
\ref{calib}. Damped precessional oscillations are clearly seen,
diminishing in amplitude (20$^{\circ}$ to 5$^{\circ}$) and
increasing in frequency (1.6 to 2.6 GHz) with increasing
longitudinal bias $0<H_{B}<100\:\textrm{Oe}$.

Single-domain Landau-Lifshitz-Gilbert (LLG) simulations of Ni
magnetization dynamics are plotted with the experimental data.
Consistent parameters are used for the fits. These are M$_{s}$=830
kA/m, measured by SQUID and $g_{eff}=2.05$, extracted from
previous PIMM measurement.  The anisotropy field
H$_{K}$=11.4$\pm$0.5 Oe, determined according to the Kittel
relationship
$\omega_{p}^{2}=\gamma^{2}\mu_{0}^{2}M_{s}H_{\parallel}$ for
in-plane magnetization precession and plotted in the right bottom
panel of Figure \ref{llg}, agrees within experimental error with
that measured statically by element-resolved hysteresis loops
(Fig. \ref{calib}, right.) The pulse field $H_{P}(t)$ was taken to
be proportional to the transmitted current waveform $i_{pulse}(t)$
at the oscilloscope; scaling for the pulse was found by fitting
the steady-state pre- and post-falling edge levels of
magnetization angle $\phi_{Ni}(0.5)$ and
$\phi_{Ni}(\textrm{4.5ns})$, respectively, yielding maximum pulse
field level $H_{P}$=7.2 Oe.

\begin{figure}[htb]
\includegraphics[width=\columnwidth]{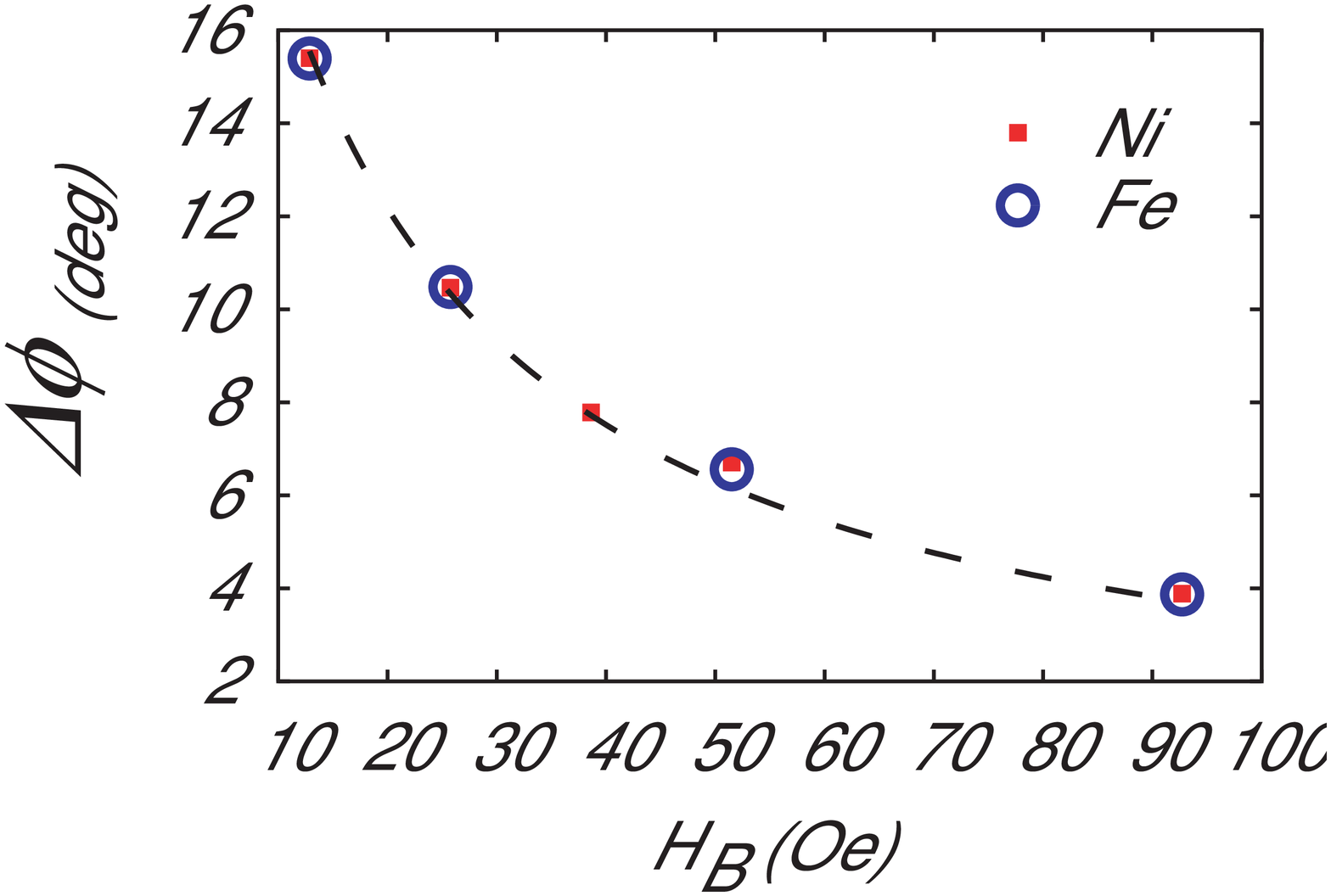}
\caption{Differences between high and low saturation angles of the
elemental Ni and Fe scans during TR-XMCD measurement, $\Delta
\phi=\phi(\textrm{t=0.5ns})-\phi(\textrm{t=4.5ns})$.  Measured
points for Ni and Fe are shown with the single-domain model
fit.\label{diffangle}}\end{figure}

The Fe magnetization angles have been processed similarly.
Differences between high and low saturation angles for Fe and Ni
TR-XMCD scans ($\Delta \phi_{Ni,Fe}$) are plotted in Figure
\ref{diffangle} as a function of longitudinal bias $H_{B}$.
Agreement between experimental $\Delta \phi_{Ni}(H_{B})$,
experimental $\Delta \phi_{Fe}(H_{B})$, and the single-domain
model is satisfactory after application of an 11\% increase in the
XMCD-to-$\phi$ calibration factor found from the Fe hysteresis
loop (Fig. \ref{calib}).  This level of adjustment is close to the
level of drift measured in the saturation level of the Fe loop.
High saturation levels of Fe are then offset to match those of Ni,
by values varying from 0-2$^{\circ}$.

\begin{figure}[htb]
\includegraphics[width=\columnwidth]{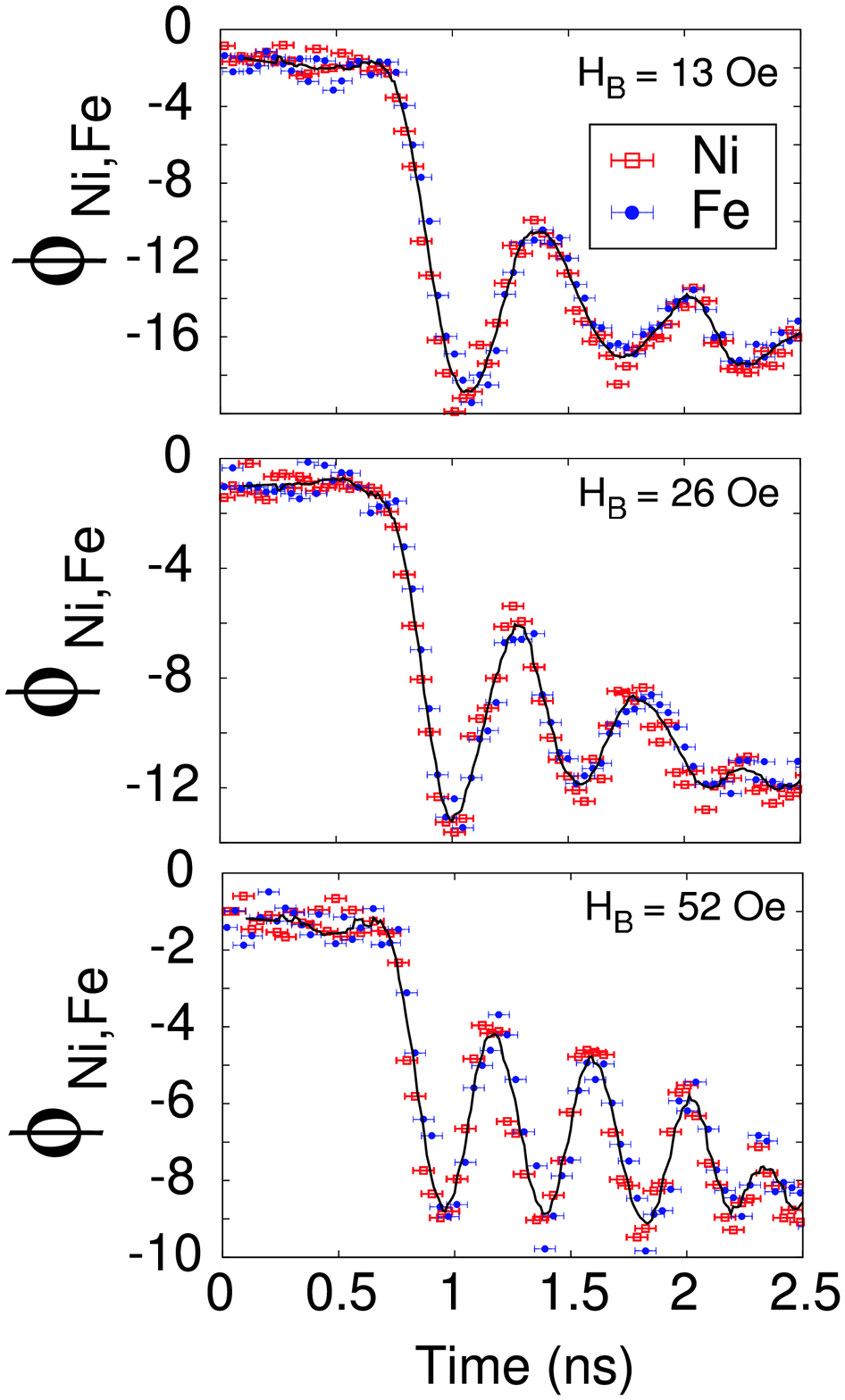}
\caption{TR-XMCD measurement of Fe and Ni magnetization angle
during free precession.  {\it Top:} $H_{B}=\textrm{13 Oe}.$ {\it
Middle:} $H_{B}=\textrm{26 Oe}.$  {\it Bottom:} $H_{B}=\textrm{52
Oe}.$ Lines are a polynomial smoothing function applied to both
data sets; timing errors of $\pm$ 45 ps are
indicated.\label{both-fe-ni}}\end{figure}

Element-specific precession is shown in Figure \ref{both-fe-ni}.
Ni and Fe moments are shown to precess together within
instrumental resolution ($\pm$45 ps).  As the longitudinal bias
field $H_{B}$ increases, the precessional frequency increases for
both elements and there is no apparent development of a phase lag
greater than experimental error.

\section{Discussion}

Ni and Fe magnetic moment precession is observed to be coupled in
phase, within 90 ps, during FMR precession of Ni$_{81}$Fe$_{19}$.
The measurement provides direct experimental confirmation that the
net Fe and Ni elemental moments are collinear during precession at
time scales exceeding 90 ps.  It should be noted that this result
could not have been easily determined without assumption from
existing measurement techniques such as PIMM or time-resolved
magnetooptical Kerr effect.

The result indicates that the exchange force exceeds the tendency
of moments on Ni and Fe to precess or relax at different rates.
The time resolution of the experiment is on the order of the
spin-lattice relaxation time $\tau_{sl}$ characterized for
Gd\cite{vaterlaus-gd}, 100 $\pm$ 80 ps, which characterizes the
direct spin-to-lattice relaxation channel; these values are
unknown for transition metals but can be expected to be within an
order of magnitude. A tendency for Fe and Ni moments to relax at
different rates is plausible based on the orbital to spin moment
ratios $L/S$ of Ni and Fe characterized in Ni$_{81}$Fe$_{19}$ by
XMCD, 0.12 and 0.08 respectively.\cite{krill}

The prospect of partially-coupled motion is more likely in systems
where exchange coupling is weaker and/or magnetic moments are more
localized. Transition-metal - rare-earth alloys, featuring lower
Curie temperatures and more isolated $f$-shells on the
rare-earths, would be attractive systems to study using the
technique. Many rare-earth dopants in Ni$_{81}$Fe$_{19}$ have been
found to increase the FMR loss\cite{bailey-apl2003}, and TR-XMCD
can test the possibility that some loss is carried by a phase lag
between RE and TM moments.\cite{clarke-t-t-yig:yt} Similarly, the
trilayer structures used in spin electronics (e.g. NiFe/Cu/CoFe)
offer the possibility of continuous adjustment of interlayer
coupling through the spacer thickness.

Finally, we note the possibility of additional studies through
more detailed spectroscopic information.  Spin and orbital moment
dynamics may be separated on individual elements through the use
of sum rules.  We take the point time resolution as that of the
measurement; it should be noted that edge resolution, with long
enough averaging and continuous monitoring for drift of the pulse
amplitude $H_{P}$, could be improved.  Better control over pulse
delivery jitter will allow a point resolution limited only by
$\sigma_{ph}$ and longer averaging could allow edge resolution
better than that of the lower bound of $\tau_{sl}$.

\section{Conclusion}

We have separated Ni and Fe moment motion in Ni$_{81}$Fe$_{19}$ by
TR-XMCD, demonstrating elemental resolution in ferromagnetic
resonance up to 2.6 GHz.  Coupled motion of elemental moments is
found to a time resolution of $\pm$ 45 ps and an angular
resolution better than 2$^{\circ}$.

\section{Acknowledgements}

We thank Gary Nitzel (NSLS) for technical support.  This work was
partially supported by the Army Research Office, grant
ARO-43986-MS-YIP.  Use of the Advanced Photon Source was supported
by the U. S. Department of Energy, Office of Science, Office of
Basic Energy Sciences, under Contract No. W-31-109-Eng-38.
Research was carried out in part at the National Synchrotron Light
Source, Brookhaven National Laboratory, which is supported by the
U.S. Department of Energy, Division of Materials Sciences and
Division of Chemical Sciences, under Contract No.
DE-AC02-98CH10886.


\end{document}